\documentclass[a4paper,11pt]{article}
\pdfoutput=1
\usepackage{graphicx}
\usepackage{bm,latexsym,amsmath,amssymb,amsfonts,mathrsfs}
\usepackage{color}
\input{colordvi.tex}

\usepackage{multirow}
\usepackage{cite}

\newcommand{\slb}{\scalebox}

\makeatletter

 \@addtoreset{equation}{section}
\makeatother

\makeatletter
\def\tbcaption{\def\@captype{table}\caption}
\def\figcaption{\def\@captype{figure}\caption}
\makeatother

\begin{document}
\allowdisplaybreaks{

\thispagestyle{empty}


\begin{flushright}
KOBE-COSMO-16-12,
TIT/HEP-656
\end{flushright}

\vspace{20mm}

\begin{center}
\slb{2.3}{Nonlocal $\mathcal{N}=1$ Supersymmetry}

\vspace{15mm}

Tetsuji Kimura\,$^{a,b}$, \
Anupam Mazumdar\,$^{c,d}$, \
Toshifumi Noumi\,$^{e,f}$ \
and \
Masahide Yamaguchi\,$^{b}$

\vspace{7mm}

\slb{.9}{\it \renewcommand{\arraystretch}{1.0}
\begin{tabular}{rl}
$a$ & Research and Education Center for Natural Sciences, Keio University, 
\\
& Hiyoshi 4-1-1, Yokohama, Kanagawa 223-8521, Japan
\\
$b$ & Department of Physics, Tokyo Institute of Technology,
\\
& Tokyo 152-8551, Japan
\\
$c$ & Consortium for Fundamental Physics, Physics Department, Lancaster University, 
\\
& LA1 4YB, UK
\\
$d$ & Kapteyn Astronomical Institute, University of Groningen, 
\\
& 9700 AV Groningen, The Netherlands
\\
$e$ & Institute for Advanced Study, Hong Kong University of Science and Technology, 
\\
& Clear Water Bay, Hong Kong
\\
$f$ & Department of Physics, Kobe University, Kobe 657-8501, Japan
\end{tabular}
}
\end{center}

\vspace{5mm}

\noindent
{\renewcommand{\arraystretch}{.9}
\begin{tabular}{r@{\!\;\;}l}
\footnotesize{Email:}&
\footnotesize{{\tt tetsuji.kimura"at"keio.jp}, \
{\tt a.mazumdar"at"lancaster.ac.uk}, \
{\tt iasnoumi"at"ust.hk},} \
\\
& \footnotesize{{\tt gucci"at"phys.titech.ac.jp}}
\end{tabular}
}

\vspace{20mm}

\begin{abstract}
We construct $\mathcal{N}=1$ supersymmetric nonlocal theories in four
dimension. We discuss higher derivative extensions of chiral and vector
superfields, and write down generic forms of K\"ahler potential and
superpotential up to quadratic order. We derive the condition in which
an auxiliary field
remains non-dynamical, and the dynamical scalars and fermions are free from the ghost degrees of freedom. We also investigate the nonlocal effects on the
supersymmetry breaking and find that supertrace (mass) formula is
significantly modified even at the tree level.
\end{abstract}

\newpage
\section{Introduction}

Supersymmetry (SUSY) is perhaps one of the most powerful extensions of
physics beyond the standard model, which attempts to unify both spin and
charge of a particle by extending the Poincar\'e group and Lie
algebra \cite{Golfand:1971iw,Haag:1974qh}. It provides an elegant
answer to the electroweak hierarchy problem by protecting the Higgs
mass, and also provides gauge couplings unification at scales close to
the grand unified scale \cite{Martin:1997ns}.

In this paper we would like to discuss higher derivative extension, especially nonlocal extension, of SUSY.
It is generally believed that higher
derivative theories can soften the ultraviolet (UV) properties. The
propagator for such theories will be more suppressed. However, even at
the classical level, the introduction of higher derivative terms in an
action is quite dangerous because there is a famous Ostrogradsky theorem
\cite{Ostrogradsky:1850fid}, which relies on having a momentum associated with higher derivative in the theory in which the energy is seen to be linear, as opposed to quadratic, states that there is 
a classical instability  
unless the theory is degenerate
\cite{Woodard:2006nt,Woodard:2015zca}.  
One way is to consider a
degenerate theory, in which the momenta associated with higher
derivative terms are not invertible. The famous example is Galileon
\cite{Nicolis:2008in}.
Supersymmetric extension of those higher derivative theories have been studied recently in \cite{Khoury:2010gb,Gama:2011ws,Khoury:2011da,Nitta:2014fca, Addazi:2015dxa, Aoki:2015eba}.

Another way to circumvent Ostrogradsky ghost is to consider infinitely higher
derivative theory (nonlocal theory),
where no such highest momentum operator can be readily identified, nor there are any extra poles in the propagator which could correspond to new degrees of freedom, such as ghosts or otherwise.
Moreover, it has been known that infinite derivatives would definitely improve the ultraviolet properties of the theory.
In particular a nonlocal extension of the Einstein
gravity has a variety of interesting properties and applications (see, e.g.,
\cite{Pais:1950za, Stelle:1976gc,Tomboulis:1980bs,Tomboulis:1983sz,Moffat:1990jj,Tomboulis:1997gg,Biswas:2005qr,Barnaby:2007ve,Barnaby:2008tc,Modesto:2011kw,Biswas:2011ar,Biswas:2014yia,Talaganis:2014ida,Tomboulis:2015gfa,Biswas:2016egy,Biswas:2016etb,Talaganis:2016ovm,Pius:2016jsl}).
It is also known that nonlocal theories capture certain aspects of string theory, particularly in the context of string field theory and p-adic string (see, e.g.,~\cite{Okawa:2012ica,Calcagni:2013eua,Taylor:2003gn,Ohmori:2001am,Moeller:2002vx,Pius:2016jsl,Freund:1987kt,Freund:1987ck,Brekke:1988dg,Ghoshal:2000dd,Minahan:2001pd}).
Nonlocal field theories would therefore be useful for constructing and understanding UV complete (gravitational) theories.

Based on such backgrounds, we wish to incorporate SUSY in nonlocal field theories. In this paper we discuss the matter and gauge field sector in particular (see the recent paper~\cite{Giaccari:2016kzy} for the gravitational sector).
Typically, in the off-shell
formalism of SUSY construction, an auxiliary field is introduced to
balance the degrees of freedom between bosons and fermions. Then, one
may wonder what should be the condition we may require in order to keep
the auxiliary field non-dynamical, when infinite derivatives are
introduced, and how it is related to the condition for the absence of a
ghost or tachyons in physical fields.

These important questions must be addressed in order to
construct a viable nonlocal SUSY theory. Phenomenologically, it is an
interesting question to ask; how the supertrace (mass) formula gets modified. 
In a global SUSY model, the supertrace (mass) formula
vanishes even after the SUSY breaking, albeit radiative corrections
slightly modifies it, which implies that not only heavier superpartners
but also lighter ones must appear.

In this paper, first of all, we will construct ${\cal N}=1$ infinitely
higher derivative extensions of chiral (neutral) superfields up to
quadratic orders in four dimensions.  We will clarify the condition how
to keep the auxiliary field non-dynamical and the absence of
ghosts. Then, we extend our construction to vector superfields including
charged chiral superfields. Finally, as a simple example of the SUSY
breaking, we shall consider a nonlocal extension of O'Raifeartaigh
model and discuss how the supertrace formula is modified. Finally,
conclusions and discussions will be given.

\section{Higher derivative action for chiral superfields}

In this section, we would like to introduce a higher derivative extension
of the standard SUSY action for chiral superfields. Let us consider,
\begin{align}
& S=
\int d^4xd^4\theta
 \,K(\Phi_i,\Phi_i^\dagger,D_\alpha,\bar{D}_{\dot{\alpha}},\partial_\mu) 
+\left[
\int d^4xd^2\theta \, W(\Phi_i,\Phi_i^\dagger,D_\alpha,\bar{D}_{\dot{\alpha}},\partial_\mu)
+ \text{h.c.}
\right]
\,, \label{higher_derivative_ansatz}
\end{align}
where the K\"ahler potential $K$, and the superpotential $W$,
constructed from $\Phi_i$'s, $\Phi_i^\dagger$'s, and their derivatives
are vector and chiral superfields, respectively. This makes the action
\eqref{higher_derivative_ansatz} SUSY because super
transformations of D-terms and F-terms are total derivatives\footnote{We
follow the notation of Wess and Bagger's book \cite{Wess:1992cp} in this paper.}.
In the following we shall construct a higher derivative action of the form
\eqref{higher_derivative_ansatz} up to the second order in $\Phi_i$ and
$\Phi_i^\dagger$, and introduce a SUSY nonlocal field theory.

\subsection{Higher derivative extension of K\"ahler potential}

We begin with the higher derivative extension $K$, of the K\"ahler
potential.  Since we just require the reality condition $K^\dagger=K$
to preserve SUSY, it is straightforward to write down the
concrete form of $K$. 

Ingredients for the second order action can be
classified into the following two: (1) One contains one chiral and one
anti-chiral superfields, and (2)  The other contains terms with two chiral
superfields and their Hermitian conjugates. A general form of the
quadratic action with one chiral superfield, $\Phi_i$, and one
anti-chiral superfield, $\Phi_i^\dagger$, is given by\footnote{Terms of
the form
$\sigma^\mu_{\alpha\dot{\alpha}}\,D^\alpha\Phi_i\bar{D}^{\dot{\alpha}}\partial_\mu\Phi_j^\dagger$
and $D^2\Phi_i\bar{D}^2\Phi_j^\dagger$ can be reduced
to~\eqref{PhiPhi^dagger} by integrating by parts.  Also, e.g.,
$D^2\Phi_i \Phi_j^\dagger$ vanishes after integration.}
\begin{align}
\label{PhiPhi^dagger}
\int d^4xd^4\theta \left[
\Phi_i\,f_{ij}(\Box)\Phi_j^\dagger
+\text{ h.c. }
\right]\,,
\end{align}
which can be thought of as a higher derivative extension of kinetic terms\footnote{Note that there is an implicit scale,   $f_{ij}(\Box/M)$, where $M$ is the scale of nonlocality. The local
two derivative theory can be attained, i.e. $f_{ij} (\Box/M^2)\rightarrow 1$ by taking the limit, $M \rightarrow \infty$. In order to avoid cluttering our formulae, we will suppress $M$.}.
In terms of component fields, it can be written as
\begin{align}
\int d^4xd^4\theta \Big[ \Phi_i\,f_{ij}(\Box)\Phi_j^\dagger
+\text{h.c.}
\Big]
&= 
\int d^4x \Big[
\phi_if_{ij}(\Box)\Box\phi_j^*
+F_if_{ij}(\Box)F_j^*
-i\psi_i f_{ij}(\Box)\sigma^\mu\partial_\mu\bar{\psi}_j
+\text{h.c.} \Big]
\,,
\end{align}
where our notation for component fields is following:
\begin{align}\label{comp}
&\Phi_i=\phi_i(y)+\sqrt{2}\theta\phi_i(y)+\theta^2 F_i(y)
\qquad
{\rm with}
\quad
y^\mu=x^\mu+i\theta \sigma^\mu\bar{\theta}\,.
\end{align}
Similarly, terms with two chiral superfields and their conjugates are
generally of the form\footnote{ Note that $\int d^4xd^4\theta\Phi_i^2$
vanishes for example.}:
\begin{align}
\int d^4xd^4\theta
\Big[
\Phi_ig_{ij}(\Box)D^2\Phi_j
+\text{h.c.}
\Big]
&= 
-4\int d^4xd^2\theta\, \Big[ 
\Phi_ig_{ij}(\Box)\Box\Phi_j \Big] 
+\text{h.c.}
\,. \label{PhiPhi}
\end{align}
The above term can be thought of as a higher derivative extension of the mass
term after integrating by parts. The above equation can be recast in terms of 
the components, \eqref{comp},
\begin{align}
\int d^4xd^4\theta \,\Phi_ig_{ij}(\Box)D^2\Phi_j+\text{h.c.} 
&=
-4\int d^4x \, \Big[ \phi_ig_{ij}(\Box)\Box F_j+F_ig_{ij}(\Box)\Box \phi_j-\psi_ig_{ij}(\Box)\Box \psi_j \Big]
\,.
\end{align}
To summarize, the higher derivative extension of the K\"ahler potential now
leads to two types of second order action; higher derivative extension of
kinetic term and mass term.  

In principle extending our analysis beyond quadratic in superfield to  third and higher order
will be straightforward, though algebraic calculations become more
complicated as we go beyond quadratic order in superfield.

\subsection{Higher derivative extension of superpotential}

Next we consider higher derivative extension of the superpotential, $W$.
Compared to the K\"ahler potential, the construction of $W$ is
rather complicated, because we require the chiral condition
$\bar{D}_{\dot{\alpha}}W=0$ to preserve SUSY. 

We can solve this
condition explicitly at the second order level in $\Phi_i$ and
$\Phi_j^\dagger$, and show that all higher derivative terms in the
superpotential can be absorbed into the K\"ahler potential and do not
generate new operators.  

As a result, a general form of higher derivative quadratic action is
given by
\begin{align}
\nonumber
S&=\int d^4xd^4\theta \Big[
\Phi_i\,f_{ij}(\Box)\Phi_j^\dagger
+\Phi_i\,g_{ij}(\Box)D^2\Phi_j
\Big]
+ \int d^4xd^2\theta \, \bar{m}_{ij}\Phi_i\Phi_j
+\text{h.c.}
\\
&=
\int d^4xd^4\theta\,
\Phi_i\,f_{ij}(\Box)\Phi_j^\dagger
+\int d^4xd^2\theta \, \Phi_i m_{ij}(\Box)\Phi_j
+\text{h.c.}
\,,
\end{align}
where $m_{ij}(\Box)=-4g_{ij}(\Box)\Box+\bar{m}_{ij}$
can be thought of as the higher derivative extension of the mass term.

\subsection{Second order action and physical spectra}

We now discuss the physical spectrum of higher derivative quadratic
action in the following generic form:
\begin{align}
\label{quadratic_chiral}
S_2
&=
\int d^4xd^4\theta\,
\Phi_if_i(\Box)\Phi_i^\dagger
+\left[\int d^4xd^2\theta\,\Phi_im_{ij}(\Box)\Phi_j
+\text{h.c.}\right]\,,
\end{align}
where $f_i$'s are real functions of d'Alembertian
and $m_{ij}$'s are complex symmetric functions $m_{ij}=m_{ji}$.

Also note that we diagonalized the (higher derivative extension of) kinetic terms.
In terms of component fields,
it can be written as
\begin{align}
S_2 &=
\int d^4x \, \Big[
\phi_i\,f_i(\Box)\Box\phi_i^*
+F_i\,f_i(\Box)F_i^*
-i\psi_i f_i(\Box)\sigma^\mu\partial_\mu\bar{\psi}_i
\nonumber \\
&\qquad \qquad
+ \Big( \phi_im_{ij}(\Box)F_j-\frac{1}{2}\psi_im_{ij}(\Box)\psi_j
+\text{h.c.} \Big)
\Big]
\,.
\end{align}
An important point here is that the auxiliary fields, $F_i$'s, acquire the
kinetic term for a general choice of $f_i$'s.  The scalars, $\phi_i$'s,
and the fermions, $\psi_i$'s, also obtain additional dynamical degrees
of freedom in general. 

\subsection{Dynamical degrees of freedom}

Now, we need to understand the true dynamical degrees of freedom - in order to 
clarify under what conditions dynamical degrees of freedom would be the 
same as that of the standard local theory, in the limit when $f_{i}(\Box) \rightarrow 1$, 
we need to first complete the square with respect to $F_i$'s:
\begin{align}
\nonumber
&S_2=\int d^4x \, \Big[
\phi_i\,f_i(\Box)\Box\phi_i^*
-\phi_jm_{ij}(\Box)\,f_i(\Box)^{-1}m_{ik}^*(\Box)\phi_k^*
\\
\nonumber
&\qquad \qquad \qquad
-i\psi_i f_i(\Box)\sigma^\mu\partial_\mu\bar{\psi}_i
-\frac{1}{2}\left(\psi_im_{ij}(\Box)\psi_j
+\text{h.c.}\right)
\\
& \qquad \qquad \qquad
+\left(F_i+f_i(\Box)^{-1}m^*_{ij}(\Box)\phi^*_j\right)\,f_i(\Box)\left(F_i^*+f_i(\Box)^{-1}m_{ik}(\Box)\phi_k\right)
\Big]\,.
\end{align}
Note that in order to keep $F_i$'s auxiliary, or non-dynamical fields,
$f_i(\Box)$'s must have {\it no zeros}, equivalently, $f_i^{-1}(\Box)$'s
must have {\it no poles}. It should be noticed that, at this stage,
the positivity of $f_i(\Box)$'s is not necessarily required.

For simplicity, let us assume that $m_{ij}(\Box)$ is diagonal and real:
$m_{ij}(\Box)=\delta_{ij}m_i(\Box)$ and $m_i(\Box)^*=m_i(\Box)$.
The action after integrating out the auxiliary fields $F_i$'s is then given by:
\begin{align}
S_2&=\int d^4x \Big[
\phi_i\, f_i(\Box)\left(\Box +f_i(\Box)^{-2}m_i(\Box)^2\right)\phi_i^*
\nonumber
\\
&\qquad \qquad 
-i\psi_i f_i(\Box)\sigma^\mu\partial_\mu\bar{\psi}_i
-\Big( \frac{1}{2}\psi_im_i(\Box)\psi_j +\text{h.c.} \Big)
\Big]
\,.
\end{align}
The on-shell conditions for $\phi_i$ and $\psi_i$ are then given by the equation of motion:
\begin{align}
f_{i} (\Box) \left( \Box +f_i(\Box)^{-2}m_i(\Box)^2 \right)\phi_{i}=0\,, \nonumber\\
f_{i} (\Box)^{2} \left( \Box +f_i(\Box)^{-2}m_i(\Box)^2 \right)\psi_{i}=0\,.
\end{align}
Here, we would like to discuss the true dynamical degrees of freedom participating in 
any classical dynamics. Now, if we demand that this infinite derivative theory maintains 
the original degrees of freedom corresponding to that of a local $2$-derivative theory, then
we need the following conditions:

\begin{itemize}

\item{{\bf $f_i(\Box)$'s must not contain any zeroes}: This is required in order to maintain $F_i$'s
non-dynamical degrees of freedom.}

\item{{\bf At most $1$-zero from $\left( \Box +f_i(\Box)^{-2}m_i(\Box)^2 \right)$}: Since $f_i(\Box)$'s
do not contain any zero, therefore $\left( \Box +f_i(\Box)^{-2}m_i(\Box)^2 \right) =0$
should have only one solution for the $\Box$. All of the other cases
lead to additional degree of freedom.}

\item{{\bf $f_{i}(\Box)>0$}: In addition, if we
require that this dynamical degree of freedom has healthy kinetic term
(that is, correct signature), $f_i(\Box)$'s must be positive. Otherwise,
this dynamical degree of freedom itself becomes ghost.}

\end{itemize}

 These conditions
are satisfied only when $f_i(\Box)$'s (or equivalently
$f_i^{-1}(\Box)$'s) is {\it exponential of an entire function},
i.e. $e^{-\gamma(\Box)}$, where $\gamma(\Box)$ is an entire function, such a function does not introduce any pole in the complex plane.
For $\gamma>0$, as $\Box\rightarrow \infty$, it is easy to see why the propagator is even more convergent in the UV.

In our case, one simple choice which would reproduce the original local spectrum could be
$m_i(\Box)=\bar{m}_if_i(\Box)$ with $\bar{m}_i$ being the mass in the
local theory, and $f_i(\Box)\sim e^{-\gamma(\Box)}$.

\section{Introducing gauge sector}

In this section we will introduce a vector
superfield by gauging the covariant derivatives.

\subsection{Gauge covariant derivatives}

Let us  consider an Abelian gauge symmetry, an extension to
non-Abelian case will be straightforward. We will define general superfields
with the charge $(p,q)$ by the following transformation rule,
\begin{align}
\mathcal{O}_{p,q}\to\mathcal{O}'_{p,q}=e^{ip\Lambda}e^{-iq\Lambda^\dagger}\mathcal{O}_{p,q}\,.
\end{align}
Note that the complex conjugate of the operator $\mathcal{O}_{p,q}$ has a charge $(q,p)$ in our convention. 
The gauge covariant extension of the spinorial derivatives, $D_\alpha$ and $\bar{D}_{\dot\alpha}$, is then defined by
\begin{align}
&\mathcal{D}_\alpha\mathcal{O}_{p,q}=D_\alpha\mathcal{O}_{p,q}+p\left(D_\alpha
 V\right)\mathcal{O}_{p,q}
\,,  \\
&\bar{\mathcal{D}}_{\dot\alpha}\mathcal{O}_{p,q}=\bar{D}_{\dot\alpha}\mathcal{O}_{p,q}+q\left(\bar{D}_{\dot\alpha} V\right)\mathcal{O}_{p,q}\,,
\end{align}
where the vector superfield, $V$, transforms as $V\to
V+i\left(\Lambda^\dagger-\Lambda\right)$.
We also introduce the vectorial gauge covariant derivative, as
\begin{align}
\mathcal{D}_\mu\mathcal{O}_{p,q}
\ &= \ 
-\frac{i}{4}\bar{\sigma}{}_\mu^{\dot{\alpha}\alpha}\{\mathcal{D}_\alpha,\bar{\mathcal{D}}_{\dot\alpha}\}\mathcal{O}_{p,q}
\ = \ 
\partial_\mu\mathcal{O}_{p,q}+pB_\mu \mathcal{O}_{p,q}+q\tilde{B}_\mu\mathcal{O}_{p,q}\,,
\end{align}
where $B_\mu$ and $\bar{B}_\mu$ are defined by
\begin{align}
B_\mu=-\frac{i}{4}\bar{\sigma}{}_\mu^{\dot{\alpha}\alpha}\left(\bar{D}_{\dot\alpha}D_\alpha V\right)\,,
\quad
\tilde{B}_\mu=-\frac{i}{4}\bar{\sigma}{}_\mu^{\dot{\alpha}\alpha}\left(D_\alpha\bar{D}_{\dot\alpha}V\right)\,,
\end{align}
with the following gauge transformations,
\begin{align}
B_\mu\to B_\mu-i\partial_\mu\Lambda\,,
\quad
\tilde{B}_\mu\to \tilde{B}_\mu+i\partial_\mu\Lambda^\dagger\,.
\end{align}
It should be noticed that the vector covariant derivative,
$\mathcal{D}_\mu$, does not commute with $\mathcal{D}_\alpha$, and
$\mathcal{D}_{\dot\alpha}$, which suggest that the vector
covariant derivative of chiral superfields do not satisfy the chirality
condition:
\begin{align}
\left[\bar{D}_{\dot\alpha},\mathcal{D}_\mu\right]\Phi=q\left(\bar{D}_{\dot\alpha}B_\mu\right)\Phi=\frac{i}{2} q W^\alpha\sigma_{\mu\alpha\dot\alpha}\Phi\,,
\end{align}
where $W_\alpha$ is the gauge invariant field strength, defined later.

It is then straightforward to gauge covariantize the matter sector by
using the covariant derivatives introduced above. For example, the
general quadratic action~\eqref{quadratic_chiral} for the chiral
superfields $\Phi_1$ and $\Phi_2$ with the charges $(1,0)$ and $(-1,0)$,
respectively, is simply covariantized as
\begin{align}
S&=
\int d^4xd^4\theta\,
\Big[
\Phi_1^\dagger\,e^{+gV} f_{1}(\mathcal{D}_\mu^2) \Phi_1 
+ \Phi_2^\dagger\,e^{-gV} f_{2}(\mathcal{D}_\mu^2) \Phi_2 
\Big]
\nonumber \\
&\quad
+\int d^4xd^2\theta \, 
\Phi_1 \bar{D}_{\dot\alpha} \bar{D}^{\dot\alpha}
\mathcal{D}^{\alpha}\mathcal{D}_{\alpha}
g(\mathcal{D}_\mu^2)\Phi_2
+\text{h.c.}
\,,
\end{align}
up to terms with field strengths. 
We should note that the
covariantization of a given matter theory is not unique, because we may
always add terms with field strengths such as $(\partial_\mu B_\nu-\partial_\nu B_\mu)^2\Phi^\dagger\Phi$ and $W_\alpha^2\Phi^2$.
The extensions of the matter sector are summarized in
Table \ref{table:extension}.
\begin{center}
\slb{1}{${\renewcommand{\arraystretch}{1.35}
\begin{array}{c||c|c|c} \hline
& \text{canonical} & \text{second order} & \text{higher order} 
\\ \hline\hline
\multirow{2}{*}{neutral}
& \big[ \Phi_i^{\dagger} \Phi_i \big]_{\text{D}} 
& \big[ \Phi_i^{\dagger} \Box \Phi_i \big]_{\text{D}} 
& \big[ \Phi_i^{\dagger} f_{ij} (\Box) \Phi_j \big]_{\text{D}}
\\
& \big[ \Phi_i \Phi_j \big]_{\text{D}} 
& \big[ \Phi_i D_{\alpha} ^2 \Phi_j \big]_{\text{D}} \simeq \big[ \Phi_i \Box \Phi_j \big]_{\text{F}} 
& \big[ \Phi_i \Box g_{ij} (\Box) \Phi_j \big]_{\text{F}} 
\\ \hline
\multirow{3}{*}{charged}
& \big[ |\Phi_1|^2 \, e^{+g V} \big]_{\text{D}} 
& \big[ \Phi_1^{\dagger} \, e^{+gV} {\cal D}_{\mu}^2 \Phi_1 \big]_{\text{D}}
& \big[ \Phi_1^{\dagger} \, e^{+gV} f_1({\cal D}_{\mu}^2) \Phi_1 \big]_{\text{D}}
\\
& \big[ |\Phi_2|^2 \, e^{-g V} \big]_{\text{D}} 
& \big[ \Phi_2^{\dagger} \, e^{-gV} {\cal D}_{\mu}^2 \Phi_2 \big]_{\text{D}} 
& \big[ \Phi_2^{\dagger} \, e^{-gV} f_2({\cal D}_{\mu}^2) \Phi_2 \big]_{\text{D}}
\\
& \big[ \Phi_1 \Phi_2 \big]_{\text{D}} 
& \big[ \Phi_1 {\cal D}_{\alpha}^2 \Phi_2 \big]_{\text{D}}  \simeq \big[ \Phi_1 \bar{D}{}_{\dot{\alpha}}^2 {\cal D}_{\alpha}^2 \Phi_2 \big]_{\text{F}}
& \big[ \Phi_1 \bar{D}{}_{\dot{\alpha}}^2 {\cal D}_{\alpha}^2 \, g({\cal D}_{\mu}^2) \Phi_2 \big]_{\text{F}} 
\\ \hline
\end{array}
}$}
\tbcaption{Higher derivative extensions of the quadratics of the neutral and charged chiral superfields in the superfield formalism. 
Here $[\ldots]_{\text{D}}$ and $[\ldots]_{\text{F}}$ are abbreviation of the integral $\int d^4 \theta$ and $\int d^2 \theta$, respectively. Also $\simeq$ here implies that the both sides are the same up to total derivative terms and an overall coefficient.}
\label{table:extension}
\end{center}
The first column implies the chiral superfields $\Phi_i$ are neutral or charged under the gauge symmetry.
In the second column the canonical term in each chiral superfield is described.
The third column gives a certain extension of each canonical term involving the second order derivatives.
In the fourth column the terms involves higher order derivatives provided by the functions $f_{ij} (\Box)$ and $g_{ij} (\Box)$ for the neutral chiral superfields, and $f_1 ({\cal D}_{\mu}^2)$, $f_2 ({\cal D}_{\mu}^2)$, and $g ({\cal D}_{\mu}^2)$ for the charged chiral superfields.
Under a theoretical constraint such as anomaly free, there exists a certain relation among their functions.
We emphasize that $\bar{D}{}_{\dot{\alpha}}^2$ in $[\ldots]_{\text{F}}$ is the chiral projection operator acting on a non-chiral operator ${\cal D}_{\alpha}^2 \Phi_i$.

\subsection{Gauge sector}

We now briefly discuss the gauge field. The field strength is encoded in gauge invariant superfields, defined by
\begin{align}
W_\alpha=-\frac{1}{4}\bar{D}^2D_\alpha V\,,
\quad
\bar{W}_{\dot{\alpha}}=-\frac{1}{4}D^2\bar{D}_{\dot{\alpha}} V\,.
\end{align}
Since $W_\alpha$ and $\bar{W}_{\dot{\alpha}}$ are chiral and antichiral
superfields, respectively. We can show that a general quadratic
action constructed from $W_\alpha$ and $\bar{W}_{\dot\alpha}$ is given
by
\begin{align}
S_{W}&=
\frac{1}{4}\left[\int d^4xd^2\theta\,W^\alpha g(\Box)W_\alpha
+\text{h.c.}
\right]\,.
\end{align}
Note that $\int d^4xd^4\theta\, W^\alpha
\sigma^\mu_{\alpha\dot{\alpha}}f(\Box)\partial_\mu\bar{W}^{\dot\alpha}$
and higher derivative extensions of the Fayet-Iliopolous (FI) term are total
derivatives.

\section{Infinite derivative extension of O'Raifeartaigh model}

Before we conclude, let us discuss  briefly an infinite derivative extension of the O'Raifeartaigh model,
where we shall discuss how the mass formula get modified.

Let us begin with the following action:
\begin{align}
S&=\int d^4xd^4\theta\,
\Phi_if_i(\Box)\Phi_i^\dagger
+\left[\int d^4xd^2\theta\,\Phi_1m(\Box)\Phi_2
+\lambda\Phi_0
+g\Phi_0\Phi_1\Phi_1
+\text{h.c.}\right]\,.
\end{align}
For the moment, we leave $f_i(\Box)$ and $m(\Box)$
as arbitrary functions of d'Alembertian
satisfying $f_i(0)=1$ and $m(0)=m$, in order to reach the local limit in the IR.
Note that we keep the cubic interactions local for simplicity.
Also there are no derivative terms for linear terms
because they are total derivatives.
In terms of component fields,
the action is given by:
\begin{align}
\nonumber
S&=\int \!d^4x\bigg[
\sum_i\left(\phi_if_i(\Box)\Box \phi_i^*
+F_if_i(\Box)F_i^*
-i\psi_if_i(\Box)\sigma^\mu\partial_\mu\bar{\psi}_i\right)
\\
&\qquad \qquad
+\left[\lambda F_0+\left(F_1m(\Box)\phi_2+F_2m(\Box)\phi_1-\psi_1m(\Box)\psi_2\right)
\nonumber \right. \\
&\qquad \qquad 
\left.
+g\left(F_0\phi_1^2+2F_1\phi_0\phi_1
-\psi_1\psi_1\phi_0-2\psi_0\psi_1\phi_1\right)\right]+\text{h.c.}
\bigg]\,.
\end{align}
By integrating out the auxiliary fields $F_i$,
we obtain the action of the form:
\begin{align}
\nonumber
S&=\int d^4x\bigg[
\sum_i\left(\phi_if_i(\Box)\Box\phi_i^*
-i\psi_if_i(\Box)\sigma^\mu\partial_\mu\bar{\psi}_i\right)
\\
\nonumber
&\qquad\qquad
-\psi_1m(\Box)\psi_2-\bar\psi_1m(\Box)\bar\psi_2
-\lambda
-\lambda g(\phi_1^2+\phi_1^{*2})
\\
&\qquad\qquad
-\phi_2 f_1(\Box)^{-1}m(\Box)^2\phi_2^*
-\phi_1f_2(\Box)^{-1}m(\Box)^2\phi_1^*+\ldots
\bigg]\,,
\end{align}
where the dots stand for cubic and quartic terms in $\phi_i$.
Since homogeneous equations of motion
are the same as that of the original (local) O'Raifeartaigh model, in our case, 
we obtain two classes of vacua, which we shall discuss below.

Here let us consider the spectrum
around the vacuum $\phi_0=\phi_1=\phi_2 =0$.
The on-shell condition for $\phi_0$ and $\psi_0$
is given by
\begin{align}
f_0(\Box)^2\Box=0\,,
\end{align}
and that for $\psi_1$, $\psi_2$, and $\phi_2$ is given by
\begin{align}
f_1(\Box)f_2(\Box)\Box-m(\Box)^2=0\,.
\end{align}
On the other hand,
the $\phi_1$ sector is modified by the $\lambda$ interaction, as
\begin{align}
f_1(\Box)f_2(\Box)\Box-\left(m(\Box)^2\pm2\lambda g\right)=0\,,
\end{align}
where the plus/minus sign is for the real/imaginary part of $\phi_1$.
Let us then choose $f_i(\Box)$ and $m(\Box)$ as the following Gaussian
operator\footnote{First of all, $f_i(\Box)$ and $m(\Box)$ need not be
the same. For simple illustration, we have chosen them to be the same,
so that the spectrum of fermions, $\phi_0$, and $\phi_2$ are the same as
the local case.
Also, we have a choice for the sign in the exponential
factor. In the plus case,
for the large $\lambda g/M^2$, the propagator
of $\sigma$ does not have a physical pole instead of $\pi$, but, in this
case, the vacuum $\sigma$ has an tachyonic potential and is
unstable.
Note that we
could select a large class of entire function instead of the Gaussian operator in principle, for instance
see \cite{Edholm:2016hbt}.}
\begin{align}
f_i(\Box)=m(\Box)=e^{-\Box/M^2}\,,
\end{align}
where we have explicitly introduced the scale of nonlocality, $M$.
If we decompose $\phi_1$ into the real  and
the imaginary part,  as $\phi_1=\frac{1}{\sqrt2}(\pi+i\sigma)$,
the masses of $\pi$ and $\sigma$ are affected by the nonlocality. The on-shell
conditions then give;
\begin{align}
e^{-\Box/M^2}(\Box-m^2)=\pm2\lambda g\,,
\end{align}
and if we
introduce a dimensionless parameter $x=\Box/M^2$, the above
equation can be expressed as
\begin{align}
\label{on-shell_O'Raiferataigh}
e^{-x}\left(x-\frac{m^2}{M^2}\right)=\pm \frac{2\lambda g}{M^2}\,.
\end{align}
Plotting the behavior of (\ref{on-shell_O'Raiferataigh}) in Figure \ref{Fig:spectraO'Raiferataigh},
we can read off a couple of interesting phenomena.
\begin{center}
\includegraphics[width=80mm, bb=0 0 360 214]{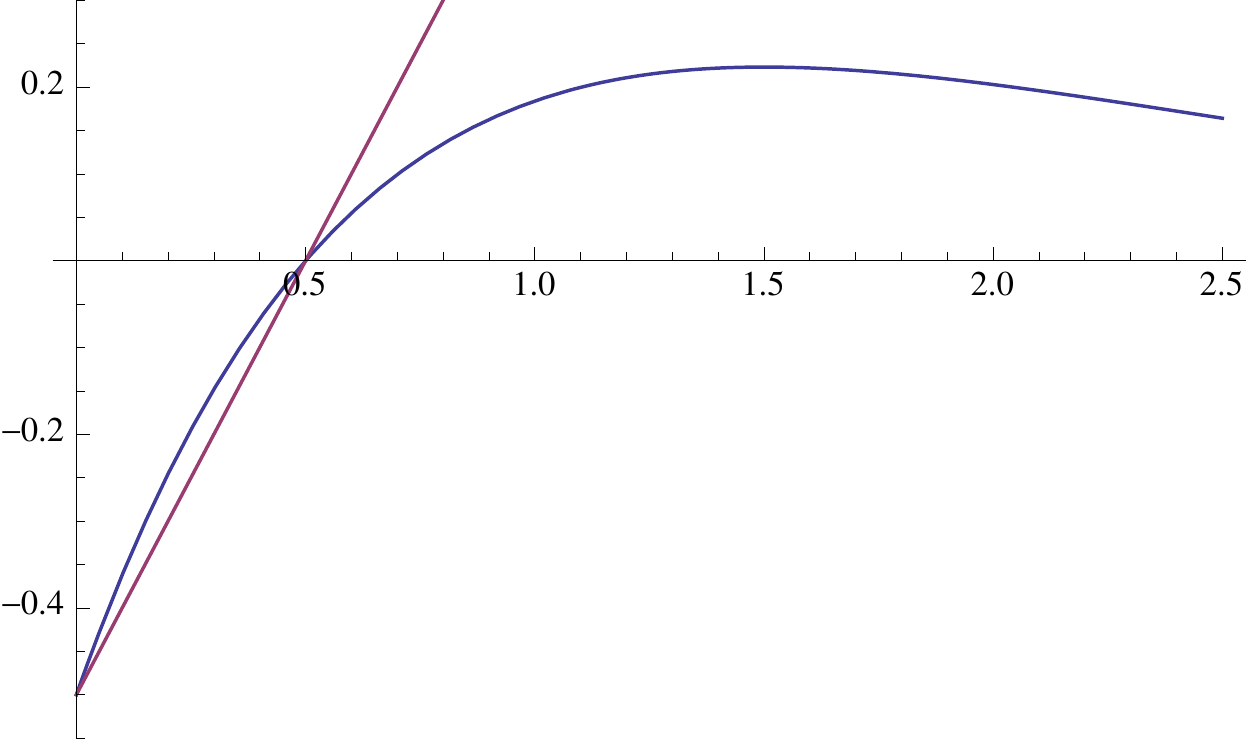}
\figcaption{The straight line denotes  $(x-\frac{m^2}{M^2})$ without any nonlocality, while 
the curved line indicates the operator $e^{-x}(x-\frac{m^2}{M^2})$ in (\ref{on-shell_O'Raiferataigh}) which involves the nonlocal effect by the factor $e^{-x}$.
Here we set $m^2/M^2=0.5$.}
\label{Fig:spectraO'Raiferataigh}
\end{center}
In this figure, a numerical value of $2 \lambda g/M^2$ in the right-hand side of (\ref{on-shell_O'Raiferataigh}) is represented as a horizontal line.
It turns out that the propagator of $\pi$ does not have a physical pole
for $2\lambda g/M^2 \gtrsim 0.22$. 
This is because the horizontal line expressing $2 \lambda g/M^2$ does not cross the curved line beyond that value. In other words there does not exist a solution of (\ref{on-shell_O'Raiferataigh}). This situation is analogous to the open string tachyon condensation in the level truncated theory (see, e.g.,~\cite{Ellwood:2001py}).

If $2\lambda g/M^2$ takes a value $0<2\lambda g/M^2\lesssim0.22$, the horizontal line cross the curved line at two points.
This means that $\pi$ has two poles.
However, one of the poles has a wrong sign, i.e., the pole gives rise to the negative norm and it provides an unphysical mode.
Such a parameter region should be avoided in order for our simple nonlocal model to be trustable at the UV scale. We would like to emphasize that such a dangerous signal appears even at the tree-level.\footnote{
In nonlocal theories the ghost degrees of freedom may easily arise in such a condensation phase unless we carefully define the theory. It will be interesting to explore under which conditions such dangerous ghosts may be avoided.
For example, string theory, which is protected by a large symmetry, will be useful to explore this direction.}

On the other hand, 
the situation for $\sigma$ is not so different from the local case (when
$2\lambda g>0$). 
Furthermore, in the limit $M\rightarrow \infty$, the mass formula reduces to that of the local case.

\section{Conclusions and discussions}

In this paper, we have constructed $\mathcal{N}=1$ supersymmetric
nonlocal theories in four dimension. We discuss higher derivative
extensions of chiral and vector superfields, and write down generic
forms of K\"ahler potential and superpotential up to quadratic order. We
find that there are only nonlocal extensions of the standard canonical
kinetic term and the mass term. Based on this action, we derive the
condition for (neutral) chiral superfields in which an auxiliary field
remains non-dynamical, and find that the same condition is necessary to
remove the ghost degree of freedom from dynamical fields. The extension
to charged chiral fields are straightforward and the complete treatment
will be discussed in the future publication \cite{KMNY}. We have also
investigated the nonlocal effects on the supersymmetry breaking. As a
concrete example, a nonlocal extension of O'Raiferataigh model is
discussed. The on-shell condition for each field is derived and we find
that that supertrace (mass) formula is significantly modified even at
the tree level, which has interesting implications on collider physics
and cosmology.

\section*{Acknowledgments}
T.K. is supported in part by the MEXT-Supported Program for the
Strategic Research Foundation at Private Universities ``Topological
Science'' No.\ S1511006, and the Iwanami-Fujukai Foundation.
A.M. is supported by the Lancaster-Manchester-Sheffield Consortium for
Fundamental Physics under STFC grant ST/J000418/1. The work of T.N. is supported by a grant from Research Grant
s Council of the Hong Kong
Special Administrative Region [HKUST4/CRF/13G].  This work
is supported in part by the JSPS Grant-in-Aid for Scientific Research
Nos.~25287054 (M.Y.) and 26610062 (M.Y.) and MEXT Grant-in-Aid for
Scientific Research on Innovative Areas ``Nuclear Matter in Neutron
Stars Investigated by Experiments and Astronomical Observations'' No.\
15H00841 (T.K.) and ``Cosmic Acceleration'' No.\ 15H05888 (M.Y.).


\end{document}